\def \kev{\rm{keV}}

\def \cm{~\rm{cm}}
\def \s{~\rm{s}}
\def \km{~\rm{km}}

\def \K{~\rm{K}}
\def \g{~\rm{g}}

\def \AU{~\rm{AU}}

\def \yr{~\rm{yr}}

\documentclass[12pt,preprint]{aastex}

\shorttitle{Eta Carinae Binarity}
\shortauthors{Soker}
\slugcomment{Draft version of \today}

\begin{document}

\title{ACCRETION BY THE SECONDARY IN ETA CARINAE DURING THE SPECTROSCOPIC EVENT:
\newline
I. FLOW PARAMETERS}

\author{Noam Soker}
\altaffiltext{1}
{Department of Physics, Technion$-$Israel Institute of Technology,
Haifa 32000 Israel;
soker@physics.technion.ac.il}

\begin{abstract}

We examine the influence of the gravity of the companion
(the secondary) to the massive primary star $\eta$ Carinae
on the winds blown by the primary and the secondary.
The two winds collide with each other after passing through two
respective shock waves, and escape the system while strongly emitting
in the X-ray band.
While during most of the 5.5 years orbital period, the
companion's gravity has a negligible effect on the winds,
we find that near periastron, {{{{ the companion's gravity may significantly
influence the flow and the }}}} companion might accrete
from the primary's wind {{{{ under certain circumstances. }}}}
Near periastron passage, the collision region of the two winds {{{{ may }}}}
collapse onto the secondary star, a process that could
substantially reduce the X-ray luminosity.
We suggest that {{{{ such an }}}} accretion process {{{{ produces }}}} the
long, almost flat, X-ray minimum in $\eta$ Carinae.

\end{abstract}

\section{INTRODUCTION}
\label{sec:intro}

$\eta$ Carinae is a very massive star that underwent a twenty
years long eruption, termed the Great Eruption, about 160 years
ago (Davidson \& Humphreys 1997).
A massive nebula, $\ga 12 M_\odot$ (Smith et al.\ 2003b),
expelled during the Great Eruption, formed the
Homunculus$-$an expanding bipolar nebula around $\eta$ Car
{{{ (Morse et al. 1998);
{{{{ A smaller bipolar nebula is believed to have been created }}}}
during the Lesser eruption of 1890 as well }}} (Ishibashi et al.\ 2003).
It is widely accepted now that $\eta$ Carinae is a
massive binary system with an orbital period of $5.54 \yr$ (e.g.,
Damineli 1996; Damineli et al.\ 1997, 2000; Ishibashi et al.\ 1999;
Corcoran et al.\ 2001a,b; 2004b; Pittard \& Corcoran 2002;
Duncan \& White 2003; Fernandez Lajus et al.\ 2003;
Smith et al.\ 2004; Whitelock et al.\ 2004; Verner et al.\ 2005).
The more massive companion of the $\eta$ Car binary system
will be referred to here as the primary, while the companion,
probably an O-type star, will be referred to as the secondary.

The binary nature of $\eta$ Car is inferred from the {{{{ periodicity
of the }}}} so-called spectroscopic event$-$the fading of high excitation
lines (e.g., Damineli et al.\ 2000).
Since the spectroscopic event might in principle result from a
mass-shell ejection by the primary star (Zanella et al.\ 1984;
Davidson et al.\ 1999; Smith et al.\ 2003a;
Martin \& Koppelman 2004), it is also called a shell ejection event.
The main motivation to assume a shell ejection is in a single star model;
in the binary model the shell ejection is not necessary, although it
might occur (Soker 2005).
The same periodicity is seen in many wave bands, from the IR
(e.g., Whitelock et al.\ 2004), to the X-ray
(Corcoran et al.\ 2001a; Corcoran 2005; Corcoran et al.\ 2004a,b).
The period {{{{ derived from the spectroscopic events }}}} is $\sim 5.54 \yr$
($2023 \pm 3$ days as given by the IR lightcurve of Whitelock et al.\ 2004),
{{{{ and is believed to represent the orbital period. }}}}
{{{{ The orbit is believed to have a }}}} very high eccentricity.
For primary and secondary masses of $M_1 = 120 M_\odot$
and $M_2=30 M_\odot$, respectively,
the semi major axis of the orbit is $a = 16.6 \AU$
(there is not yet agreement on all the binary parameters, e.g.,
Ishibashi et al.\ 1999; Damineli et al.\ 2000; Corcoran et al.\ 2001a,
2004b; Hillier et al.\ 2001; Pittard \& Corcoran 2002;
Smith et al.\ 2004).

Less agreement exists concerning the role of the secondary in the shaping
of the wind from the primary in $\eta$ Car.
Based on multi-color photometry van Genderen et al.\ (1994, 1995, 1999)
proposed the presence of an accretion disk around a secondary star
before the periodicity of the spectroscopic event was discovered
(Damineli 1996).
Based on theoretical calculations Soker (2001b, 2003, 2004, 2005) argued
that a single star cannot explain many of the properties of $\eta$ Car,
in particular that a single star cannot account for the
bipolar shape of the Homunculus.
Instead, {{{{ those papers argued that }}}} the secondary accreted a
large fraction of the mass that was expelled in the Great Eruption
forming an accretion disk and two jets which
shaped the wind of the primary into the bipolar Homunculus (Soker 2001b).
Soker (2003, 2005) discusses the formation of an accretion disk
in the present day binary system, in agreement with some
of the suggestions made by van Genderen et al.\ (1994, 1995, 1999).

Soker (2005) discussed three basic accretion phases, two of which occur
in the present day.
Both stars blow winds, with the primary's wind having a much larger mass
loss rate and a lower velocity.
The winds collide, and at one location the momentum fluxes of
the two winds exactly balance each other, forming a stagnation point.
The stagnation point is located close to, but not exactly on,
the line joining the centers of the two stars
because of orbital motion.
The shocked material of the primary star cools very fast
(Pittard \& Corcoran 2002; see eq. 1 in Soker 2003), namely,
before the mass moves far from the stagnation point.
The surrounding pressure then compresses the cooled post-shock
gas to high densities.
When blowing the wind, radiation pressure on the escaping gas overcomes
gravitational attraction.
However, this might not be the case with the dense gas near the
stagnation point.
Gravitational force on the dense and slowly moving (much
below escape velocity) gas there might
become large enough to accrete part of the mass back onto
one of the stars.
Whether accretion occurs at all, and if it does, which of the
two stars accrete most of the gas depends on the accretion phase of
$\eta$ Car:

\begin{enumerate}

\item {\bf The Great Eruption.}
The mass loss rate by the primary was very high, and the
stagnation point was within the secondary's
Bondi-Hoyle accretion radius.
Along the entire orbit the
secondary steadily accreted mass with high specific angular momentum.
An accretion disk was formed and two jets (or a collimated fast wind$-$CFW)
were launched (Soker 2001b).
{{{{ According to Soker (2001b) }}}} these jets shaped the two lobes which
are now observed as the Homunculus.
\item {\bf Apastron passages.}
Soker (2003) proposed that during present apastron
passages the primary itself can accrete $\sim 5\%$ of the mass
lost over an entire orbit.
This should not be confused with the accretion fraction of
$\sim 50 \%$ by the secondary during the 20 year span of the
Great Eruption.
The high specific angular momentum of the accreted gas implies the
formation of an accretion disk around the primary.
The primary star might blow a CFW.
\item {\bf Periastron passages.}
Near periastron passages, which occur near the spectroscopic events,
short accretion episodes might occur, possibly leading to pulsed
ejection of two jets by the secondary (Soker 2005).
This accretion process was proposed but not studied by Soker (2005).
The secondary might also ionize a non-negligible region in its
surrounding neighborhood.
\end{enumerate}

This paper presents further exploration of the nature of the wind interaction
near periastron passage.
In the next section we compare the typical time scales and length scales
of several processes during the periastron passage.
{{{{ The reader interested only in the main points of the proposed model
can skip section 2 and go directly to section 3, where }}}}
we discuss how these {{{{ may allow }}}} an accretion
event onto the secondary star for several weeks near periastron passage.
Our main results are summarized and discussed in section 4.

\section{RELEVANT LENGTH AND TIME SCALES AT PERIASTRON}
\label{sec:time}

\subsection{Length Scales}

We calculate the distance of the stagnation point of the secondary's
wind taking into account the orbital motion; by orbital velocity we
refer to the relative orbital velocity of the two stars.
Following Usov (1992), we take the wind with the larger momentum flux, in
our case the primary's wind, to be plane parallel near the secondary star.
Because of the orbital motion, the stagnation point of the secondary's
wind will be ahead of the line joining the centers of the two stars
(Figure 1).
The radial (along the line joining the two stars) component of the
relative velocity between the secondary star and the primary's wind is
$v_1-v_r$, where $v_1$ is the {{{{ radial component of the }}}} primary's
wind speed and $v_r$ the radial component of the orbital velocity;
$v_r$ is negative when the two stars approach each other.
Because the stagnation point is very close to the secondary star,
we assume that this is also the relative radial velocity of the stagnation
point to the primary's wind.
The total relative speed between the stagnation point and the
primary's wind is
\begin{equation}
v_{\rm wind1} = \left(v_\theta^2 + (v_1-v_r)^2 \right)^{1/2},
\label{vwind1}
\end{equation}
where $v_\theta$ is the tangential component of the orbital velocity.
We neglect any time delay between ejection of the wind by the primary and
its collision with the secondary's wind. This assumption
is justified over most of the orbit, but not near periastron,
where the orbital and primary's wind speed are almost equal.
However, the uncertainties because of this assumption are less problematic
than the uncertainties caused by our ignorance on the exact velocity profile
of the primary's wind, i.e., its acceleration over a distance of $\sim 3 \AU$,
and a possible enhancement in the primary's mass loss rate near periastron.
At the stagnation point the ram pressures of the two winds are equal
\begin{equation}
\rho_1 v_{\rm wind1}^2=\rho_2 v_2^2,
\label{stag1}
\end{equation}
where $v_2$ is the secondary's wind speed, assumed to be much larger
than all other flow velocities in the problem.

The respective wind densities are
\begin{equation}
\rho_i=\frac{\dot M_i}{4 \pi D_i^2 v_i}, \qquad i=1,~2,
\label{stag1b}
\end{equation}
where $D_i$ is the distance from the respective star to the
stagnation point (Figure 1).
Substituting the expressions for the winds' densities in equation
(\ref{stag1}) gives
\begin{equation}
\frac{1}{D_1^2} \frac{v_{\rm wind1}^2}{v_1^2} =  \frac{\beta^2}{D_2^{2}} ,
\label{stag2}
\end{equation}
where
\begin{equation}
\beta \equiv \left( \frac{\dot M_2 v_2}{\dot M_1 v_1} \right)^{1/2}.
\label{beta}
\end{equation}
As we see $\beta$ depends on the orbital separation because near
periastron it is assumed that the primary's wind does not yet
reach its terminal speed.

Let $\phi$ be the angle measured from the secondary between the direction to the primary
and that to the stagnation point (Figure 1).
The following trigonometric relation holds
$D_1^2=D_2^2 +r^2 - 2r D_2 \cos \phi$, where $r$ is the
orbital separation, and
$\cos \phi=(v_1-v_r)/v_{\rm wind1}$.
Substituting for $D_1$ in equation (\ref{stag2}) gives
an equation which can be solved for $D_2$,
\begin{equation}
D_2= \beta r \left\{ \left[ \left( \frac{v_1-v_r}{v_{\rm wind1}}\right)^2 \beta^2
+\frac{v_{\rm wind1}^2}{v_1^2} -\beta^2 \right]^{1/2}
-\frac{v_1-v_r}{v_{\rm wind1}} \beta \right\}
\left( \frac{v_{\rm wind1}^2}{v_1^2} -\beta^2 \right)^{-1} .
\label{stag3}
\end{equation}

Near periastron the gravitational influence of the secondary on the
flow near the stagnation point should be considered.
This effect on the undisturbed primary's wind is characterized by the
Bondi-Hoyle accretion radius
\begin{equation}
R_{\rm acc2}= \frac {2 G M_2}{v_{\rm wind1}^{2}}  = 0.2   
\frac{M_2}{ 30 M_\odot} \left( \frac {v_{\rm wind1}}{500 \km \s^{-1}} \right)^{-2} AU.
\label{accrad}
\end{equation}
The relative speed $v_{\rm wind1}$ is scaled for periastron passage, where
$v_{\rm orb} \simeq 400 \km \s^{-1}$, $v_1 \sim 300 \km \s^{-1}$, since
the primary's wind did not yet reach its terminal speed of
$\sim 500 \km \s^{-1}$.
Martin et al.\ (2005), for example, take the primary's wind speed to be
$v_1 \simeq 500 (r/3 \AU) \km \s^{-1}$ for $r<3 \AU$,
and $v_1 \simeq 500 \km \s^{-1}$ for $r \ga 3 \AU$.
When approaching periastron $v_{\rm wind1}$ will be larger than its value
when leaving periastron. This results in a much larger accretion
radius when leaving periastron, as described in the next section.

As in the classical Bondi-Hoyle-Lyttleton accretion flow, the
density and velocity of the inflowing gas increase as the gas approaches the
gravitating point mass.
This increases the ram pressure of the primary's wind, and according to
equations (\ref{stag1}) {{{{ and (\ref{stag1b}) }}}} the stagnation point
distance from the secondary changes as $D_2 \propto (\rho v^2)^{-1/2}$.
Using the density (from Danby \& Camm 1957) and the velocity
(from energy conservation) along the symmetry axis in the upflow
direction (ahead of the stagnation point), one finds the stagnation
distance when the secondary's gravity is considered
\begin{equation}
D_{g2}=2 \left(1+\frac{R_{\rm acc2}}{D_2} \right)^{-1/4}
\left[1+ \left(1+\frac{R_{\rm acc2}}{D_2} \right)^{1/2} \right]^{-1} D_2.
\label{stag4}
\end{equation}
For the parameters used here, $R_{\rm acc2}/D_2 \simeq 0.5$ near periastron,
hence $D_{g2} \simeq 0.8 D_2$ near periastron.
{{{{ Some of the quantities derived in this section are plotted on Figure 2. }}}}

\subsection{Time scales}

As shown by several authors (e.g., Pittard \& Corcoran 2002; Soker 2003),
at all orbital phases the cooling time of the shocked primary's wind is much
shorter than the flow time of the gas out of the shocked region.
The post shocked primary's wind in a large area near the stagnation point cools
and is compressed by the ram pressure of the colliding winds.
As shown by Soker (2005) cold and dense blobs of size $r_b \ga 0.001 D_2$ will be
accreted by the secondary.
{{{ One of the uncertainties in this study involves the exact velocity profile
and terminal velocities of the two winds. In particular, the radiation
from one star can influence the velocity profile of the wind from the
other star before the winds collide (Gayley et al.\ 1997).
In any case, we estimate these uncertainties to be minor.
The secondary star radius is about an order of magnitude smaller than the
distance to the stagnation point.
Hence, it is already at its terminal velocity,
{{{{ assuming that the radiation pressure of the primary does not
slow down much the secondary wind.
This is justified by ($i$) the calculations of Pittard \& Corcoran (2002)
of the X-ray emission, that show the secondary wind to be shocked at
$v_2 \sim 3000 \km \s^{-1}$, with little variation in the
shocked gas temperature over most of the orbit (Ishibashi et al. 1999),
and $(ii)$ by the location of the stagnation point far from the primary. }}}}
For the primary wind we assume a simple velocity profile in section 2.3.
Taking other reasonable wind properties for $\eta$ Car from the literature will
not change much the results of this paper. }}}

{{{{ In section 3 we propose that mass from the shocked primary wind
is accreted for $\sim 80$~days near periastron passage.
This phase begins with accretion of dense blobs. }}}}
{{{ We now estimate the blobs' properties following (2005).
We consider a post-shock spherical blob of mass $m_b$, density
$\rho_b$, and radius $R_b$, located at a distance $r_2$ from the
secondary of mass $M_2$.
We make three assumptions.
$(i)$ Because of the short radiative cooling time mentioned above,
the temperature of the blob is $T_b=10^4 \K$,
\newline
$(ii)$ The blobs are in pressure equilibrium with the ram pressure
of the secondary's wind so that at any distance $r_2$,
$\rho_b k T_b/(\mu m_H)=\rho_2 v_2^2$,
where $\mu m_H$ is the mean mass per particle in the blob, and
$k$ is the Boltzmann's constant.
Using the assumed values of $\dot M_2 =10^{-5} M_\odot \yr^{-1}$
and $v_2 =3000 \km \s^{-1}$ (see section 2.3) gives
\begin{equation}
\rho_b= 5 \times 10^{-11}
\left( \frac{r_2}{1 \AU} \right)^{-2}
\g \cm^{-3}.
\label{rhob}
\end{equation}
The mass in one blob is
\begin{equation}
m_b= 7 \times 10^{-10}
\frac{\rho_b}{10^{-10} \g \cm^{-3}}
\left( \frac{R_b}{0.01 \AU } \right)^{3}
M_\odot.
\label{mb}
\end{equation}
\newline
$(iii)$ The radiative pressure of the secondary plays a minor role.
This comes from two reasons.
First, the ratio of the ram pressure of the secondary wind to
the secondary radiation pressure is
\begin{equation}
\xi \equiv \frac{P_{2{\rm rad}}}{P_{2{\rm ram}}} \simeq
\frac {L_2/c}{\dot M_2 v_2} = 0.6
\left(\frac{L_2}{9 \times 10^{5} L_\odot} \right) ,
\label{ratp}
\end{equation}
{{{{ where the secondary luminosity is scaled according to Verner et al. (2003).
Namely, the radiation pressure adds to the secondary ram pressure,
but not more than $60 \%$. }}}}
The second reason is that near the stagnation region, where it is
determined whether accretion takes place, the radiation pressure by the
primary acts in the opposite direction to that of the secondary.
As can be seen from Figure 2, the distance of the stagnation
region from the primary ($r-D_{g2}$) is $\sim 3$ times the distance to the
secondary ($D_{g2}$).
However, the primary luminosity is estimated to be
$\sim 10$ times that of the secondary.
Since the radiation pressure drops as $1/({\rm distance})^2$,
we find that the radiation pressure of the primary, which pushes the gas
toward the secondary, more than compensates for the secondary radiation
pressure near the stagnation point.
This holds as long as $L_1 \ga 10 L_2$.
Considering radiation pressure by both stars will actually increase the
effects studied in the present paper.

The condition for accretion, therefore, is that the gravitational force
\begin{equation}
f_g = \frac{G M_2 m_b}{r_2^2}
\label{fgrav}
\end{equation}
on a blob be larger than the force due to ram and
radiation pressure of the secondary wind
\begin{equation}
f_{w2} = \rho_{w2} v^2_{2} \pi R_b^2(1+\xi).
\label{fw2}
\end{equation}
Substituting the typical physical values used here and equation
({\ref{rhob}) in the condition for accretion ${f_g}> {f_{w2}}$, gives
the constraint on the size of the accreted blob
\begin{equation}
R_b  >0.004 (1+\xi) \left( \frac{r_2}{1 \AU} \right)^{2}
\AU.
\label{rbmin}
\end{equation}
This shows that even very small blobs can be accreted.
Substituting values in the last equation, we find $R_b > 0.007 \AU$ for
$r_2=D_{g2}=1 \AU$, when the orbital separation is $r \sim 4 \AU$
(accretion is not expected to start earlier than this time)
{{{{ According to the last equation closer to the secondary even smaller blobs
can be accreted.
By equations (\ref{rbmin}), and (\ref{mb}) with (\ref{rhob}), the minimum mass
allowed for accreted blobs goes as $m_b \propto r_2^{4}$.
Namely, as a blob moves toward the secondary, even if it breaks to smaller
blobs, these may still be accreted if they are not too small.
The blobs near the stagnation point can be somewhat smaller than the
constraint given by equation (\ref{rbmin}) because the radiation pressure
of the primary pushes toward the secondary.
}}}}

The last point should be emphasized.
If the flow structure was such that the primary wind streams undisturbed
toward the secondary star, to the point where the secondary radiation
pressure is larger than the primary radiation pressure
$L_2/r_2^2 > L_1/(r-r_2)^2$, then the radiation pressure could have slowed
down or could even expel the incoming primary wind; a process termed
radiative braking (e.g., Gayley et al.\ 1997).
However, the primary wind encounters a shock wave {\it before} radiative
braking starts.
Dense blobs are likely to form in this unstable interaction region.
We conclude that the radiation pressure and ram pressure cannot prevent
the accretion by the secondary of even small blobs. }}}}

As discussed in section \ref{sec:accretion}, if the stagnation region collapses
the maximum steady state accretion rate is
$\dot M_{\rm acc2} \sim 10^{-6} M_\odot \yr^{-1}$.
When the stagnation region collapses
{{{{ before or near periastron, the accretion radius is smaller and }}}}
the Bondi-Hoyle formula will give a lower accretion rate.
However, {{{{ if (as proposed in the next section) }}}}
the entire stagnation region collapses, we can still crudely take an
accretion rate of $\sim 10^{-6} M_\odot \yr^{-1}$.
As shown below the accretion dynamical time scale is $\sim 0.01 \yr$.
Using the typical blob mass from equation (\ref{mb}), we find that
$\sim 10$ blobs exist during the collapse of the stagnation point region.
These blobs when accreted onto the secondary {\it are assumed } to
vigorously disturbed the acceleration region of the secondary wind,
such that practically the secondary wind ceases to exist.
For the rest of the flat minimum phase, we assume that there is no wind
collision and the accretion flow is that of the Bondi-Hoyle-Lyttleton
type.
{{{{ That the accreted primary wind almost completely shuts down the
secondary wind is a strong assumption of the present paper, and must
be checked in future calculations of the acceleration zones of accreting
O stars, and observations of $\eta$ car during minimum. }}}}

The free fall time from the stagnation point to the secondary is
\begin{equation}
\tau_{\rm ff2}= 1.05  
\left( \frac{M_2}{30 M_\odot}\right)^{-1/2}
\left( \frac {D_{g2}}{0.2 \AU} \right)^{3/2} ~{\rm day}.
\label{freef}
\end{equation}
The outflow time from the stagnation point is somewhat longer than the
flow of the undisturbed primary wind
\begin{equation}
\tau_{\rm flow1}  \ga \tau_{\rm f1} \equiv \frac{D_{g2}}{v_{\rm wind1}} = 0.69 
\frac {D_{g2}}{0.2 \AU}
\left( \frac {v_{\rm wind1}}{500 \km \s^{-1}} \right)^{-1} ~ {\rm day}.
\label{flow}
\end{equation}

The cooling time of the secondary's wind is longer than the flow time
$\tau_{\rm flow2}$; for $v_2=3000 \km \s^{-1}$
$\tau_{\rm flow2} > \tau_{f2} \equiv 0.1 (D_{g2}/0.2 \AU) ~{\rm days}$.
However, near periastron when the stagnation distance $D_{g2}$ becomes
small, the cooling of gas very close to the stagnation point becomes
non-negligible because very close to the stagnation
point $\tau_{\rm flow2} \gg \tau_{f2}$ (Usov 1991).
The secondary's wind is shocked to a temperature of $\sim 10^8 \K$, where
the cooling function dependance on temperature is
$\Lambda =\Lambda_0 T^{0.4}$.
The cooling time at constant pressure is
$\tau_{\rm cool2} = (5/2) nkT(n_e n_p \Lambda)^{-1}$, where $n$, $n_e$
and $n_p$ are the total, electron, and proton densities, respectively.
Substituting the numerical values gives for the cooling time of the
shocked secondary's wind
\begin{equation}
\tau_{\rm cool2}=1.1
\left( \frac{\dot M_2}{10^{-5} M_\odot \yr^{-1}} \right)^{-1}
\left( \frac {D_{g2}}{0.2 \AU} \right)^{2}
\left( \frac{v_2} {3000 \km \s^{-1}} \right)^{2.2} ~ {\rm day}.
\label{tauc}
\end{equation}
(Near periastron {{{{ and for these parameters
the post shock electron density is $\sim 10^{10} \cm^{-3}$ and the temperature
$\sim 10^8 \K$; the equalization time of ion and electron temperature for these
values is $\sim 0.1$~day (Usov 1002), }}}} shorter than the cooling time.)
The fraction of the shocked wind that cools to very low temperatures
and is compressed is (eq. 7 of Usov 1991)
$\alpha \sim  (\tau_{f2}/\tau_{\rm cool2})^2 $
which is $\sim 0.01$ near periastron passage.
This implies that at periastron passage the shocked secondary's wind
within a distance of $\sim 0.1 D_{g2}$ from the stagnation
point cools and is compressed.
This will further reduce the support to the shocked primary's wind against
being accreted by the secondary.

\subsection {Numerical Values}

The parameters used in the present calculations are based
on papers cited in previous sections (e.g., Pittard \& Corcoran 2002;
Martin et al.\ 2005).
We take the primary's wind to have a profile of
$v_1=500 [1-(0.4 \AU/r_1)] \km \s^{-1}$,
where $r_1$ is the distance from the center of the primary;
as this expression is a crude estimate of the acceleration
zone of the primary's wind, we can take $r_1=r$ at the stagnation point.
At periastron $r=1.66 \AU$ and $v_1=380 \km \s^{-1}$, larger than the
speed assumed by Martin et al.\ (2005).
{{{{ Based on the results of Pittard \& Corcoran (2002) }}}}
the secondary's wind speed is taken to be $v_2=3000 \km \s^{-1}$,
and the mass loss rates are {{{{ assumed to be }}}} $\dot M_1=3 \times 10^{-4} M_\odot \yr^{-1}$
and $\dot M_2 =10^{-5} M_\odot \yr^{-1}$.
The masses are $M_1=120 M_\odot$ $M_2=30 M_\odot$
{{{{ (Hillier et al.\ 2001), }}}}
the eccentricity is $e=0.9$ {{{{ (Smith et al.\ 2004), }}}}
 and orbital period 2024 days, hence the semi-major axis is $a=16.64 \AU$.

\section{ACCRETION NEAR PERIASTRON PASSAGE}
\label{sec:accretion}

The relevant length scales and time scales derived in the previous section are
plotted as function of the orbital phase, where phase zero is taken at
periastron, in Figure 2.
Note that phase zero corresponds to periastron passage.
This is not to be confused with phase zero defined from observations of the
intensities of different lines; in the latter cases phase zero is
assumed to be near periastron, but may not correspond precisely to periastron
passage (see footnote 4 in Martin et al.\ 2005).

Two ratios determine the importance of the secondary's gravity:
the accretion radius to stagnation distance $R_{\rm acc2}/D_{g2}$, and
the flow time of the shocked primary's wind to the free fall time
from the stagnation point $\tau_{\rm f1}/\tau_{\rm ff2}$.
Over most of the orbital motion these ratios are very small
(lower panel of Figure 2), and gravity is negligible.
However, very close to periastron these ratios become $\sim 0.5$.
For example, both these ratios are larger than $\sim 0.25$ from 10 days
before to 40 days after periastron passage.
These large ratios suggest that accretion might take place
near periastron passage.
The asymmetry of these ratios around periastron fit well
with the asymmetrical behavior of the event around periastron.
For example, the X-ray emission after the flat minimum period does not
return to its luminosity prior to the flat minimum period.
{{{{ This asymmetry results from the larger relative wind velocity
$v_{wind1}$ as the system approaches periastron and $v_r<0$,
than when the system leaves periastron and $v_r>0$  (eq. {\ref{vwind1}).
This then influences the values of the stagnation distance
(eq. \ref{stag4}) and the accretion radius (eq. \ref{accrad}),
which determines the other properties, such as cooling time. }}}}

The exact flow structure requires 3D gas-dynamical simulations.
However, we can suggest the following scenario already from the
present results.
Due to thermal instabilities {{{{ (e.g., Stevens et al. 1992) }}}}
dense large blobs are formed in
the post-shock primary's wind region near the stagnation point.
These blobs are pulled to the secondary as periastron is approached.
Very close, possibly $\sim 10$~day prior, to periastron passage
the mass of the primary's wind that is accreted is
{{{{ assumed to be }}}} large enough to shut down the secondary wind.
{{{{ The assumed shut-down must be non-linear
because the mass accretion rate is smaller than the mass loss
rate of the secondary. }}}}
As the secondary's wind no longer reaches the previous stagnation region,
the entire primary's wind entering the Bondi-Hoyle accretion cylinder,
i.e. having an impact parameter smaller than the accretion radius,
will be accreted by the secondary.
In other words, the previously colliding winds region collapses
onto the secondary.
{{{{ We emphasize that a key ingredient in the model is that blobs
accreted near periastron passage shut down, or substantially weaken, the
secondary wind, such that the accretion radius becomes larger than
the stagnation distance.
This allows more accretion that is {\it assumed} to shut down the
secondary wind in several days.
This process might occur even if during the time the first blobs are accreted
the accretion radius is smaller than the stagnation distance. }}}}

In a steady state situation the mass accretion rate is
$\dot M_{\rm acc2} = \pi R^2_{\rm acc2} \rho_1 v_{\rm wind1}$, which
for the parameters used here reaches a maximum value of
$\dot M_{\rm acc2} \sim 0.006 \dot M_1 \simeq 10^{-6} M_\odot \yr^{-1}$.
This is not a high accretion rate, compared with the mass blown by
the secondary.
However, a short time after the region near the stagnation point
collapses this mass falls onto the secondary from one direction.
As shown here and by Soker (2005) dense blobs can be accreted by the secondary.
After the collapse of the stagnation point, the primary's wind accelerates
and its density increases as it approaches the secondary.
{{{{ It is assumed that the accreted cold gas prevents the
secondary's normal wind acceleration.
As the stars separate, the mass accretion rate then declines and the secondary
wind reappear again,
building the wind collision region, only when orbital separation increases
again to several AU. }}}}
The orbital separation is twice its periastron values (1.66 \AU) after 20 days,
and five times its periastron distance after 75 days.
When the colliding winds region is built again, the X-rays reappear.

In a steady state accretion from a wind, the accreted mass
possesses angular momentum.
If the specific angular momentum is larger than that of a test
particle performing Keplerian motion on the equator of the accreting
star, an accretion disk might be formed.
The condition for that is given by, e.g., Soker (2001a; eq. 1 there).
Substituting the physical parameters used here, we find that
in a steady state accretion the accreted primary's wind possesses
specific angular momentum which is too low by a factor of $\sim 10$
to form an accretion disk around the secondary.
One of the reasons is that in the Bondi-Hoyle type accretion
flow, the accretion flow rearranged itself to accrete
a small fraction ($\sim 20 \%$) of the angular momentum entering the
Bondi-Hoyle accretion cylinder.
However, during the brief collapse of the material near the
stagnation point this reduction does not happen.
Furthermore, the collision region near the stagnation point
is larger than the accretion radius, and the specific angular
momentum is larger.
Over all, we suggest that an accretion disk does form for a brief
period, a few days, during the collapse stage of the colliding wind
region (the collapse of the stagnation point).
Such an accretion disk could blow a collimated fast wind.

\section{DISCUSSION AND SUMMARY}
\label{sec:event}

In the previous section we have shown that near periastron the
secondary's gravity becomes a significant factor in determining
the flow near the stagnation point.
Along most of the orbit the accretion radius
($R_{\rm acc2}$; eq. {\ref{accrad}}),
which characterized the influence of the secondary on the
undisturbed primary wind, is much smaller than the distance of
the stagnation point from the secondary
($D_{g2}$; eq. {\ref{stag4}}).
Along most of the orbit, the outflow time of the
post-shock primary's wind from the stagnation point vicinity
($\tau_{\rm f1}$; eq. {\ref{flow}}) is much shorter than the
free fall time of this gas to the secondary
($\tau_{\rm ff2}$ ; eq. {\ref{freef}}).
However, very close to periastron passage these two ratios,
$R_{\rm acc2}/D_{g2}$ and $\tau_{\rm f1}/\tau_{\rm ff2}$
(lower row of figure 2), increase to $\sim 0.5$.
This shows that the secondary significantly alters the flow such
that it might accrete from the cool post-shock primary's wind.
{{{{ We speculate that }}}} this could lead to the collapse of the
wind collision region near the stagnation point.
{{{{ We further speculate }}}} that for a short time an accretion
disk and a collimated fast wind might be formed.
We note that the accretion phase is relatively short, and
the steady state accretion rate relatively low
($\dot M_{\rm acc2} \sim 10^{-6} M_\odot \yr^{-1}$; sec. \ref{sec:accretion}).
Therefore, the total accreted mass is negligible compared with the mass
lost in the secondary wind.

{{{{ In addition to the assumptions made in developing the model,
their are some not well determined binary parameters which
introduce further uncertainties.
We consider the results with regards to uncertainties in winds parameters
quite robust. This is because the wind parameters are constrained by the
X-ray properties (Pittard \& Corcoran 2002).
More than that, it is possible that the equatorial mass loss by the primary
will increase near periastron passages (Pittard \& Corcoran 2002;
but see Soker 2005), hence enhancing accretion.
The model is only slightly more sensitive to the eccentricity
as long as $e \ga 0.8$.
If $e=0.8$ instead of $e=0.9$, for example, we find that the maximum ratio
of $R_{\rm acc2}/D_{g2}$ is $0.23$ instead of $0.47$ for $e=0.9$.
However, in the case $e=0.9$ assumed in this paper, this ratio stays at a
value of $R_{\rm acc2}/D_{g2} > 0.2$ for $\sim 65$~days, about the length
of the flat X-ray minimum.
It is assumed here that for such a value the mass accretion rate is
high enough to shut down the secondary wind.
In the case of $e=0.8$ the inequality $R_{\rm acc2}/D_{g2} > 0.2$ holds for
$\sim 40$~ days, but the inequality $R_{\rm acc2}/D_{g2} > 0.17$ holds
for $\sim 65$~ days.
We see that the conditions for $\sim 65$~days accretion is not much different in
the two cases.
This suggests than even for $e=0.8$ accretion might occur, and our
model can hold even for eccentricity as low as $e=0.8$.
Of course, this depends on the validity of our assumptions that
($i$) accretion starts for these values of $R_{\rm acc2}/D_{g2}$;
and ($ii$) the accretion can shut down the secondary wind.
}}}}

The suggestion of the collapse of the interacting winds region
is disputed.
Other researchers argue that the winds collision region
(the stagnation point region) continues to exist during the X-ray
minimum, e.g., Abraham et al.\ (2005) who based their claim
on their suggestion that the 7~mm emission comes from this region.

In our proposed scenario the collapse of the wind collision region
to the secondary is behind the long and almost flat minimum in the
X-ray emission, lasting $\sim 60$~days (0.03 of the cycle)
(Ishibashi et al.\ 1999; Corcoran et al.\ 2001a;
Corcoran et al.\ 2004b; Corcoran 2005).
The X-ray emission results from the collision of the two winds from
the two stars (Corcoran et al.\ 2001a; Pittard \& Corcoran 2002), as in
similar massive binary systems (Usov 1992), most prominently, the
WR-O binary system WR 140 (Williams et al.\ 1990).
In the interacting massive binary system WR 140 the X-ray minimum is
not flat, and it can easily be explained by absorption of the X-ray
emission by the dense wind of the WR star
(Corcoran et al.\ 2004b; Pollock et al.\ 2005).
In $\eta$ Car, the flat minimum in X-ray emission seems to be intrinsic
(Corcoran et al.\ 2000; Hamaguchi et al.\ 2005), that is, a reduction in
emission measure of the X-ray emitting gas, and the minimum is not easy to
explain only by absorbing material.
{{{ The hard ($\ga 5 \kev$) X-ray emission drops by a factor of up to
$\sim 100$ during the minimum, but there is still hard X-ray emission.
In Akashi et al. (2006) this residual X-ray emission during minimum is
explained in the frame of the accretion model.

After the collapse of the stagnation region, the primary's wind collides
with the secondary's wind very close to the secondary, most likely
within the acceleration zone of the secondary's wind.
In this type of wind interaction the secondary's luminosity slows down the
primary's wind before it encounters the shock wave {{{ (Gayley et al. 1997),}}}
reducing substantially the X-ray intensity and making it softer (Usov 1992).
{{{{ The slowing down process by the secondary radiation pressure affects the
location of the new stagnation point, but does not push it back to its
original place.
Hence, it cannot prevent the proposed collapse of the colliding wind region. }}}}
{{{ In any case, this process is not relevant to $\eta$ Car because
the primary wind is too slow to contribute to the observed X-ray emission
above $1 \kev$. This is true even if the gravitational acceleration
by the secondary is taken into account (Akashi et al.\ 2006). }}}
The effects of the secondary's gravity and radiation on this asymmetrical
flow are very complicated, and are postponed for discussion to a future paper.
In addition to the reduction in the mass of the colliding primary's wind
and its deceleration, a substantial fraction of the X-rays emitted by the
shocked winds is absorbed by the dense accreted mass
{{{ (Akashi et al.\ 2006). }}}

We also speculate that the accretion process is connected to the
peak luminosity of the HeII $\lambda~4687$\AA\ line.
This peak occurs very shortly after the X-ray flux starts to decline
(Steiner \& Damineli 2004; Martin et al.\ 2005).
In a future paper we will examine the possibilities
that the He line is emitted by the cooling secondary's wind as its
cooling time $\tau_{\rm cool2}$ (eq. {\ref{tauc}}) becomes short
(figure 2, row 2), or that the He line is emitted by a
collimated outflow blown by the accreting secondary.

{{{{ We thank an anonymous referee for detail and very helpful comments
which substantially improved the presentation of the results. }}}}
This research was supported in part by a grant from the Israel
Science Foundation.

\end{document}